\documentclass[global,twocolumn]{svjour}
\usepackage{graphicx}% Include figure files
\usepackage{epsfig}
\usepackage{amsmath}
\usepackage{amssymb}
\usepackage{textcomp}
\usepackage{braket}

\usepackage{dcolumn}% Align table columns on decimal point
\usepackage{bm}% bold math

\begin{document}

\title{Interaction of a Laser with a Qubit in Thermal Motion and its Application to Robust and Efficient Readout}
\author{U.~Poschinger, A.~Walther, M.~Hettrich, F.~Ziesel, F.~Schmidt-Kaler}
\institute{Institut f\"ur Quantenphysik, Universit\"at Mainz, Staudingerweg 7, 55128 Mainz, Germany}

\date{\today}% It is always \today, today,
             %  but any date may be explicitly specified
\maketitle

\begin{abstract}
We present a detailed theoretical and experimental study on the optical control of a trapped-ion qubit subject to thermally induced fluctuations of the Rabi frequency. The coupling fluctuations are caused by thermal excitation on three harmonic oscillator modes. We develop an effective Maxwell-Boltzmann theory which leads to a replacement of several quantized oscillator modes by an effective continuous probability distribution function for the Rabi frequency. The model is experimentally verified for driving the quadrupole transition with resonant square pulses. This allows for the determination of the ion temperature with an accuracy of better than 2\% of the temperature pertaining to the Doppler cooling limit $T_{D}$ over a range from 0.5$T_{D}$ to 5$T_{D}$. The theory is then applied successfully to model experimental data for rapid adiabatic passage (RAP) pulses. We apply the model and the obtained experimental parameters to elucidate the robustness and efficiency of the RAP process by means of numerical simulations. 
\end{abstract}

%  37.10.Ty   Ion trapping
%  32.80.Qk   Coherent control of atomic interactions with photons

\section{Introduction}
Ions trapped in linear Paul traps allow for near-perfect realizations of qubits, which can be coherently controlled by driving on long-lived transitions between internal electronic states with lasers. Important applications range from quantum information and simulation experiments, quantum metrology or fundamental quantum optics. One of the most prominent challenges is given by the requirement to keep one or more ions sufficient well localized that sufficiently high fidelities of coherent control operations are attained. This is usually achieved by sideband cooling of the ions close to the motional ground state of one or more vibrational modes. For larger ion numbers, however, the number of motional modes increases such that complete ground state cooling becomes inconvenient. Furthermore, with the advent of segmented microtraps, it became possible to store an increased number of qubits and shuttle the qubits between different trap sites. This advantage is bought at the price of heating rates which can be larger by several orders of magnitude compared to conventional linear Paul traps. Furthermore, motional energy can be transferred during the shuttling operations. Thus, one has to resort to coherent control operations which are inherently robust against residual motion of the qubits. In order to characterize the dependence of the fidelity of these operations and find possible ways for improvement, a theoretical tool is required for simulating the coherent dynamics of the qubit under the consideration of multiple excited motional modes. The motion leads to fluctuations of the Rabi frequency, which represent an \textit{inhomogeneous} broadening effect when a single experimental run is considered, however as ion trap experiments are generally conducted by repeating a given control sequence and averaging over the qubit readout result, the fluctuations manifest themselves as an \textit{effective homogeneous} broadening. We have developed an effective Maxwell-Boltzmann theory which enables us to precisely and efficiently calculate the averaged dynamics of such a system subject to control laser pulses which are shaped in amplitude and frequency. This theoretical tool might find further applications for closed loop quantum control experiments with shaped laser pulses \cite{TIMONEY2008}, especially for optimized entangling gates working in the thermal regime \cite{KIRCHMAIR2009}. \\
We present a case study where we use this theoretical approach to model electron shelving dynamics on a quadrupole transition for a Doppler cooled ion. This has direct experimental relevance, as we encode the qubit information in the Zeeman split sublevels of the S$_{1/2}$ groundstate of a $^{40}$Ca$^+$ ion \cite{POSCHINGER2009}, and we employ electron shelving on the electric quadrupole transition S$_{1/2}\rightarrow$D$_{5/2}$ near 729~nm with a decay time of 1.17~s \cite{KREUTER2005} for qubit readout. Population from only one of the qubit levels is to be transferred to the metastable state, such that the occurrence of resonance fluorescence on a cycling transition is conditional on the qubit state. In order to realize a transfer process which is sufficiently robust against experimental parameter fluctuations, we make use of the rapid adiabatic transfer (RAP) process, where the transfer pulses are adiabatically switched on and off and the laser frequency is swept \textit{(chirped)} across the atomic resonance \cite{WUNDERLICH2007}. This makes the transfer process inherently robust against fluctuations and drifts of the laser frequency and power, as well as fluctuations of the Rabi frequency arising from thermal motion of the trapped ion. For the RAP readout, this method offers the possibility to make a determined trade between robustness, efficiency and duration of the process for a given amount of maximum available laser power. This manuscript is organized as follows: The Maxwell-Boltzmann theory for the laser driven qubit is developed in Sec. \ref{sec:theory}. We then give a brief account on our experimental apparatus and the used techniques in Sec. \ref{sec:apparatus}. In the following Sec. \ref{sec:expresults}, we use the Maxwell-Boltzmann theory to accurately model Rabi oscillations on the quadrupole transition which are subject to thermally induced dephasing. The temperature is inferred from the dephasing behavior. We make use of the model and the obtained temperature to reproduce experimental data for different RAP parameters (amplitude and chirp range) without free fit parameters. This allows us to use the model for the detailed numerical studies presented in Sec. \ref{sec:numerical}, where we investigate the dependence of the robustness and maximum fidelity of the RAP process on the control parameters.

\section{Effective Maxwell-Boltzmann theory for a laser driven qubit}
\label{sec:theory}
The interaction picture Hamilton operator of a coherently and resonantly driven qubit, which is harmonically confined in a trap, is given 
by 
\begin{equation}
\hat{H}_I=\tfrac{1}{2}\hbar\Omega_0\hat{\sigma}^{+}\left(1-\tfrac{\eta^2}{2}\left(2\hat{a}^{\dagger}\hat{a}+1\right)+\mathcal{O}\left(\eta^4\right)\right)+\textrm{h.c.},
\label{eq:hamil1}
\end{equation}
where $\Omega_0$ is the bare Rabi frequency, $\hat{a}^{\dagger}$ and $\hat{a}$ are the raising and lowering operators for the harmonic oscillator describing the motion in the trap at frequency $\omega$, and $\eta=\cos(\alpha)k\sqrt{\hbar/(2m\omega)}$ is the Lamb-Dicke factor. Here, $k$ is the wavenumber of the driving laser beam which makes an angle of $\alpha$ with the oscillation direction and $m$ is the ion mass.  The Hamiltonian Eq. \ref{eq:hamil1} is a second order expansion in terms of $\eta$, and terms describing off-resonant excitation of motional sidebands have been neglected, which is justified for $\eta\sqrt{\bar{n}}<1$. Depending on the laser beam geometry, the laser can couple to three oscillation modes with different frequencies, Lamb-Dicke factors and population distributions. This can be accounted for in the Hamiltonian Eq. \ref{eq:hamil1} by replacing the $\eta^2$ term by a summation over several modes. If the ion is initialized in the ground state $\ket{S}$ and exposed to a resonant square pulse of duration $t$, the probability of transferring the ion to the metastable state $\ket{D}$ is given by averaging the unitary dynamics over the vibrational modes:
\begin{equation}
P_D(t)=\sum_{\{n_i\}} \left(\prod_i p_{th}(n_i,\bar{n}_i)\right) \tfrac{1}{2}\left(1-\cos\left(\Omega_{\{n_i\}} t)\right)\right),
\label{eq:quadthermalrabiosc}
\end{equation}
where the index $i$=1,2,3 runs over the motional modes, such that a threefold summation over the vibrational quantum numbers $n_i$=0..$n_{\textrm{max}}$ has to be carried out.
If we assume that the vibrational modes are thermally occupied, the joint probability to find the ion with a set of quantum numbers $\{n_i\}$ is given by the product of the three respective Boltzmann distributions 
\begin{equation}
p_{th}(n,\bar{n})=\frac{\bar{n}^n}{(\bar{n}+1)^{n+1}},
\label{eq:thermalphondist}
\end{equation}
with the mean phonon numbers $\bar{n}_i^n=k_B T/(\hbar \omega_i)$ pertaining to the $i$-th mode oscillating at the frequency $\omega_i$. The Rabi frequency 
\begin{equation}
\Omega_{\{n_i\}}= \Omega_0 \prod_i M_{n_i}^{\textrm{car}}(\eta_i^2),
\end{equation}
is determined by the matrix elements \cite{LEIBFRIED2003}
\begin{eqnarray}
M_{n_i}^{\textrm{car}}(\eta_i^2)&=&\langle D,n_i|e^{i\eta_i(\hat{a}_i+\hat{a}_i^{\dagger}})|S,n_i\rangle \nonumber \\
&=&e^{-\eta_i^2/2}L_{n_i}^0(\eta_i^2),
\label{eq:matrixelements}
\end{eqnarray}
\begin{figure}[t!]\begin{center}
    \includegraphics[width=0.5\textwidth]{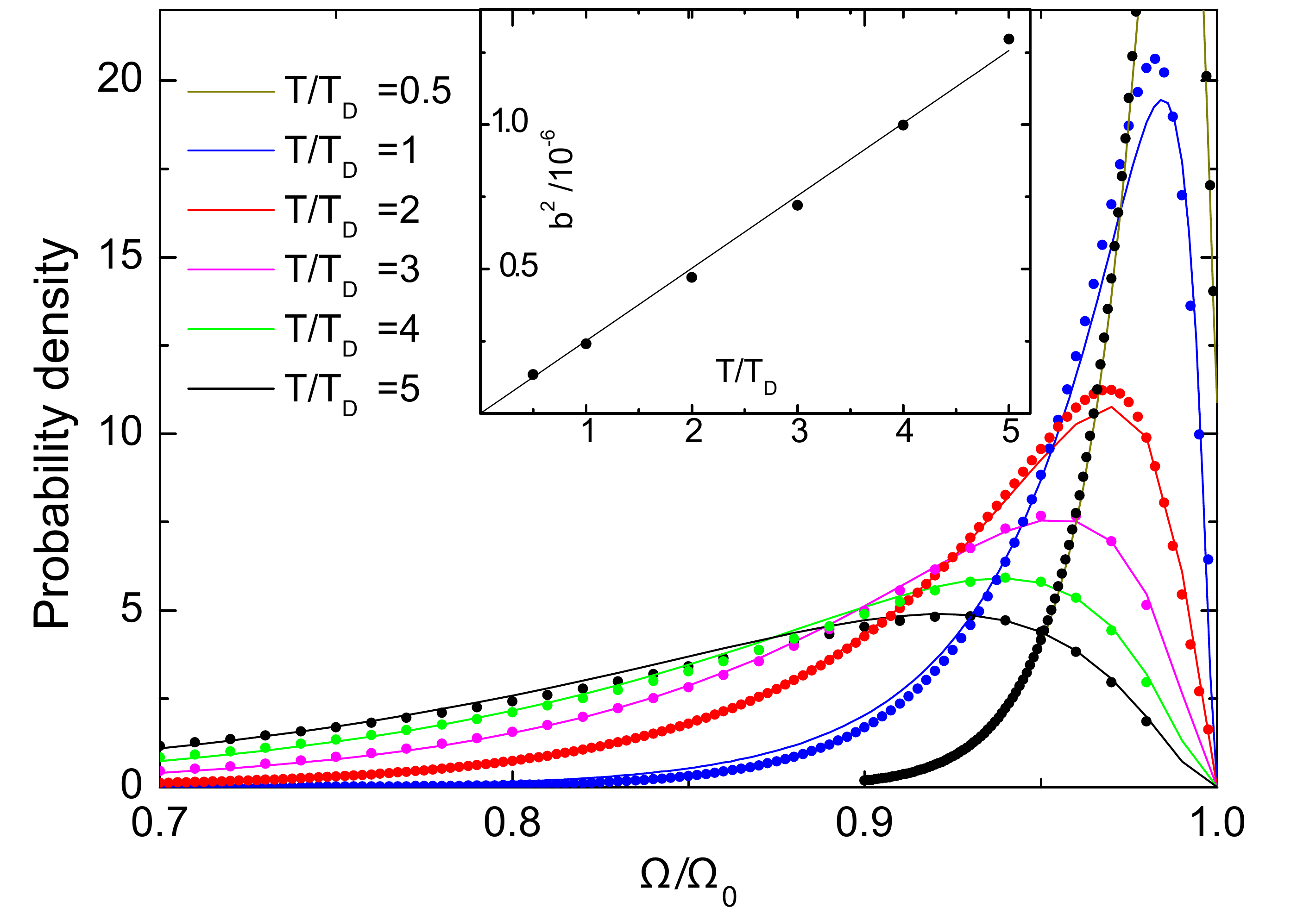}
\caption[Effective Rabi frequency distribution]{(color online) Effective Rabi frequency distribution: The symbols show sample probability densities obtained from Eq. \ref{eq:smoothedrabidist}, and the solid lines result from a fit of this data to Eq. \ref{eq:normalizedmodeledeffectiverabifreqdist}. Data for different temperatures is shown, where the temperature is given in terms of the Doppler cooling limit $T_{D}$, see text. The data set for $T\approx2\cdot T_{D}$ (red) corresponds to experimental parameters which accurately reproduce the measurement results in Fig. \ref{fig:quadcohdynamics}. The inset shows the linear relation between $b^2$ and the ion temperature $T$.}
\label{fig:effrabidist}
\end{center}\end{figure}
where the $\eta_i$ are the Lamb-Dicke parameters for mode $i$ and $L_{n}^0(\eta^2)$ is an associated Laguerre polynomial.
The triple sum in Eq. \ref{eq:quadthermalrabiosc} is numerically very inconvenient, especially if extensive simulations to arbitrary excitation pulse shapes or fitting of experimental data for ion thermometry are to be performed. Even more, in the case of multi-ion crystals, since the number of relevant vibrational modes increases linearly with the ion number, the summation approach becomes completely intractable. The problem of the time evolution on the carrier for a multi-mode thermal distribution has already been addressed in Ref. \cite{ROOS2000}, however, the approach is limited to resonant square excitation pulses. Thus, we use the alternative approach of averaging over a probability distribution function of Rabi frequencies.\\
The probability for finding a specific Rabi frequency is given by: 
\begin{equation}
w(\Omega)= \sum_{\{n_i\}}\delta\left(\Omega-\Omega_{\{n_i\}} \right)\prod_i p_{th}(n_i,\bar{n}_i).
\label{eq:threemoderabiscillation}
\end{equation}
Practically, $w(\Omega)$ is characterized by a set of Rabi frequencies $\Omega_{\{n_i\}}$ and their corresponding probabilities. These can be evaluated numerically where the motional quantum numbers are taken into account up to a truncation limit. The inaccuracy of this truncation can be estimated by summing the calculated probabilities, which should lead to a value slightly smaller than unity. We now assume that the differences between the contributing Rabi frequencies are smaller than the inverse interaction time and the individual probabilities are small due to the many contributing frequencies. Then, the frequencies can not be individually discerned and a description by a smooth continuous Rabi frequency probability distribution is justified. The probability for attaining a Rabi frequency of $\Omega$ during one individual measurement is then given by convolution of  $w(\Omega)$ with a Gaussian smoothing function: 
\begin{equation}
\tilde{w}(\Omega)=\frac{1}{\sqrt{2\pi\sigma^2}}\int_0^{+\infty}w(\Omega') e^{-\frac{(\Omega-\Omega')^2}{2\sigma^2}}d\Omega'.
\label{eq:smoothedrabidist}
\end{equation}
Under the condition that the smoothing parameter $\sigma$ is chosen to be larger than the average spacing between two Rabi frequencies and smaller than Rabi frequency differences that are resolved on experimental timescales, $\tilde{w}(\Omega)$ is empirically found to be well described by
\begin{equation}
w_b(\Omega)=\mathcal{N}\left(\frac{\Omega_0-\Omega}{\Omega}\right)^4e^{-\left(\frac{\Omega_0-\Omega}{b^2\Omega}\right)^{1/4}},
\label{eq:normalizedmodeledeffectiverabifreqdist}
\end{equation}
with the normalization factor $\mathcal{N}$, which is to be determined numerically for given parameters $b,\Omega_0$. The parameter $b$ directly determines the temperature $T$  under the assumption that the three motional modes are in thermal equilibrium. The relation between $T$ and $b$ is determined empirically by calculating $w_b(\Omega)$ for a set of $\Omega$ values and for various temperatures as shown in Fig. \ref{fig:effrabidist}. We obtain the relation
\begin{equation}
\frac{T}{T_{D}}=c\;b^2,
\end{equation}
where $c$ is uniquely determined by the specific set of Lamb-Dicke factors, i.e. by the laser geometry and the trap frequencies. The fact that the model function Eq. \ref{eq:normalizedmodeledeffectiverabifreqdist} does not perfectly match the actual probability distribution, systematic errors in the determination of the temperature occur. This leads to uncertainties in the determination of $b$ and $c$. From the numerical data shown in Fig. \ref{fig:effrabidist}, and for our specific set of Lamb-Dicke parameters given below in Sec. \ref{sec:apparatus}, we find $c=$4.0(1)$\cdot$10$^6$. An inaccuracy of the temperature of better than 0.02~T$_{D}$ over a temperature range from 0.5~T$_{D}$ to 5~T$_{D}$ can be claimed.\\
The analogy to the Maxwell-Boltzmann velocity distribution as used in kinetic gas theory becomes clear from the distributions shown in Fig. \ref{fig:effrabidist}: If more than one motional degree of freedom is taken into account, a larger density of states favors higher thermal energies, whereas the 1D Boltzmann distribution is monotonically decreasing with respect to energy, leading to a peaked probability distribution. The excess energy translates into a Rabi frequency reduction due to an average Doppler shift, broadening the optical transition. Quantitative differences to kinetic gas theory arise from the facts that i) one is dealing with a harmonically confined instead of a free particle, ii) the spatial extent of the thermal motion is comparable to the wavelength of the driving laser field (in the thermal regime), leading to nonlinear coupling matrix elements in Eq. \ref{eq:matrixelements}, and iii) the confinement anisotropy and the laser geometry lead to different coupling matrix elements for the vibrational modes. The great simplification is that instead of six parameters $\eta_i,\bar{n}_i$, the thermal motion is characterized by only one parameter, and the three-fold summation in Eq. \ref{eq:threemoderabiscillation} is replaced by a single integral:
\begin{equation}
P_D(t)=\frac{1}{2}\int_0^{\Omega_0} w_b(\Omega) \left(1-\cos(\Omega t)\right) d\Omega.
\label{eq:effectivemoderabiosc}
\end{equation}
The relation between the bare Rabi frequency $\Omega_0$ and the experimentally determined pulse time when the first excitation maximum occurs $\tau_{\textrm{max}}$ is given by:
\begin{equation}
\tau_{\textrm{max}}=\frac{\pi}{\Omega_0}(1+2^{16}b^2),
\label{eq:effectivepitime}
\end{equation}
as obtained from $dw_b/d\Omega|_{\bar{\Omega}}=0$, corresponding to the effective Rabi frequency $\bar{\Omega}=\pi/\tau_{\textrm{max}}$ with the largest probability.
The practical application of our model is as follows: \\
For a given laser-ion interaction setting (beam geometry, ion number, trap frequencies), one has to calculate the data characterizing the $w(\Omega)$ Eq. \ref{eq:threemoderabiscillation} for a set of example temperatures as shown in Fig. \ref{fig:effrabidist}. Then, the smoothed probability distribution Eq. \ref{eq:normalizedmodeledeffectiverabifreqdist} is to be determined at sample points such that the $b$ parameter can be extracted from a nonlinear regression. Once the relation between $T$ and $b$ is established and the bare Rabi frequency is calculated from Eq. \ref{eq:effectivepitime}, the model can be applied e.g. as in Eq. \ref{eq:effectivemoderabiosc} to simple Rabi oscillations.

\section{Experimental Apparatus}
\label{sec:apparatus}

In our experiment, we keep a single $^{40}$Ca$^+$ ion in a microstructured linear Paul trap \cite{SCHULZ2006,SCHULZ2008}. The oscillation frequency pertaining to the axial mode of vibration is $\omega_{\textrm{ax}}=2\pi\cdot$1.35~MHz, while the radial modes oscillate at $\omega_{\textrm{rad}}=2\pi\cdot\{$2.4,3.0$\}$~MHz. A quantizing magnetic field at 45$^{\circ}$ to the trap axis provides a Zeeman splitting of $\Delta_Z\approx$2$\pi\cdot$18~MHz between the m$_J=\pm$1/2 levels of the S$_{1/2}$ ground state, which are henceforth denoted as $\ket{\downarrow}$ and $\ket{\uparrow}$. The ion is Doppler cooled on the S$_{1/2}\rightarrow$P$_{1/2}$ cycling transition near 397~nm, while resonance fluorescence from this transition is monitored both on an EMCCD camera and a photomultiplier tube (PMT). Coherent dynamics are driven on the $S_{1/2}\rightarrow D_{5/2}$ E2 quadrupole transition near 729~nm, where laser light from an amplified diode laser system is switched and modulated by a double-pass acousto-optical modulator (AOM) running at frequencies around 80~MHz. Up to 100~mW of laser light is focused onto the trapping site with a lens of 200~mm focal length, resulting in spot size at the focus of about 30~$\mu$m FWHM. The propagation axis is aligned to be at 45$^{\circ}$ to the trap axis and orthogonal to the quantizing magnetic field. This leads to Lamb-Dicke factors for the three vibrational modes of $\{\eta_i\}=\{$0.059,0.031,0.028$\}$. The polarization of this laser is chosen to be at an angle of 45$^{\circ}$ to the plane defined by the propagation direction and the magnetic field. According to the selection rules for E2 transitions, this allows for driving of the transitions $\ket{\uparrow}\rightarrow\ket{D_{5/2},m_J=+5/2}$ and $\ket{\uparrow}\rightarrow\ket{D_{5/2},m_J=+3/2}$. The AOM is supplied with r.f. signals from a DDS-based synthesizer~\footnote{VFG-150, Toptica Photonics, Graefelfing, Germany}, which enables shaping of the laser pulses in frequency, phase and amplitude.\\
The experiments are carried out as follows: the ion is Doppler cooled on the cycling transition and initialized in $\ket{\uparrow}$ by optical pumping on the same transition. Then, the 729~nm excitation pulse is irradiated onto the ion, after which the readout is performed by irradiation on the cycling transition. If the ion was excited to the metastable state, fluorescence rates of typically about 30$\cdot$10$^3$~counts/s are observed, whereas background count rates of about 4$\cdot$10$^3$~counts/s are measured if the ion resides in the metastable state. After the readout, the ion is reset by quenching the metastable state by irradiation of laser light near 854~nm driving the D$_{5/2}\rightarrow$P$_{3/2}$ dipole transition. For a given set of excitation pulse parameters, the sequence is repeated 200 times such that the excitation probability is determined with a shot noise limited inaccuracy of at maximum 3.5\%.

\section{Experimental results}
\label{sec:expresults}
\begin{figure}[th!]\begin{center}
\includegraphics[width=0.45\textwidth]{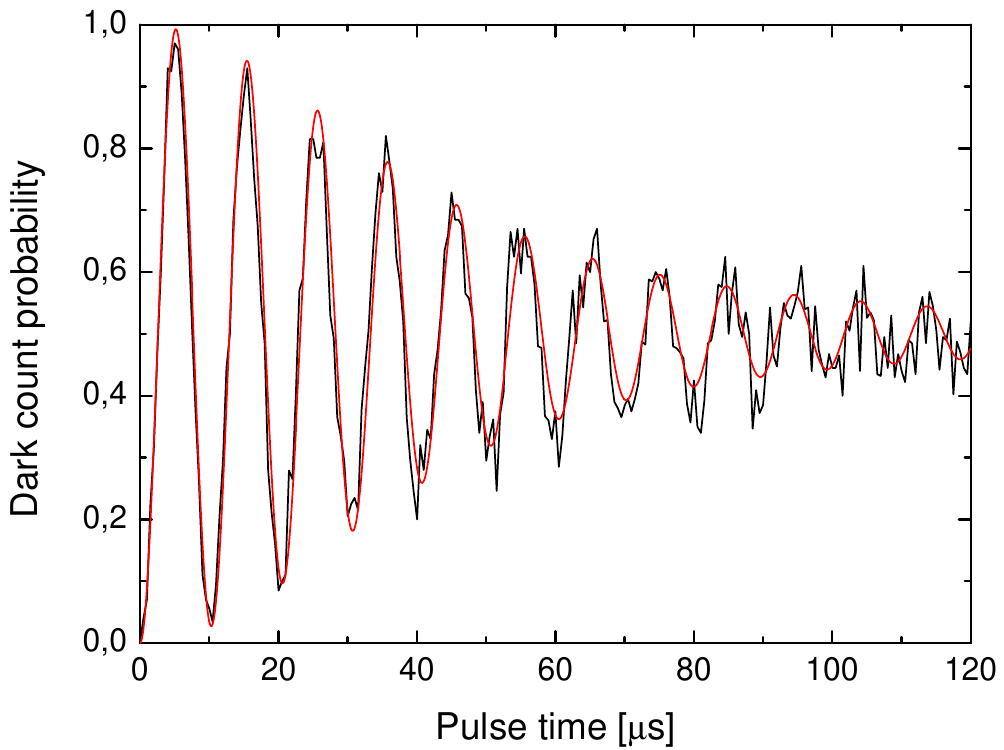}
\caption[Coherent dynamics on the quadrupole transition]{(color online) Coherent dynamics on the quadrupole transition: The fraction of population in the metastable state is plotted against the duration of a square excitation pulse with the laser frequency tuned to the $\ket{\uparrow}\rightarrow\ket{D_{5/2},m_J=+5/2}$ transition. One can clearly see a very rapid dephasing, which is due to the interaction with three thermally excited vibrational modes. Note that the dephasing behavior, i.e. the envelope of the oscillations, cannot be described by a Gaussian or exponential decay.}
\label{fig:quadcohdynamics}
\end{center}\end{figure}

The measured excitation probability on the transition $\ket{\uparrow}\rightarrow\ket{D_{5/2},m_J=+5/2}$ versus pulse duration for square pulses is shown in Fig. \ref{fig:quadcohdynamics} along with the fit to the model Eq. \ref{eq:effectivemoderabiosc}. The period allows for a direct extraction of $\tau_{\textrm{max}}$ = 4.93(5)~$\mu$s and therefore allows for the determination of the bare Rabi frequency according to Eq. \ref{eq:effectivepitime}, which is found to be $\bar{\Omega}=$2$\pi\cdot$105(1)~kHz for the shown data. With this knowledge, the parameter $b$, characterizing the distribution of the Rabi frequencies, can be fixed from a fit of the signal to the model Eq. \ref{eq:effectivemoderabiosc}. We find $b=$7.1$\cdot$10$^{-4}$, which is consistent with $T\approx 2.0\cdot T_{D}$ if we assume that all modes are cooled to the same temperature, leading to mean phonon numbers of $\{n_i\}=\{$16.9,9.5,7.6$\}$ . The temperature is found to be two times larger than the theoretical Doppler cooling limit of $T_{D}\approx$0.55~mK for an ideal two-level system. This discrepancy is attributed to a competing micromotion-induced Doppler heating effect. Further reasons for decreased Doppler cooling efficiency are the invalidity of the two-level approximation due to the presence of the Zeeman splitting (which is comparable to the linewidth of the cooling transition) and the additional decay channel to the D$_{3/2}$ state. \\
\begin{figure}[t!]\begin{center}
\includegraphics[width=0.45\textwidth]{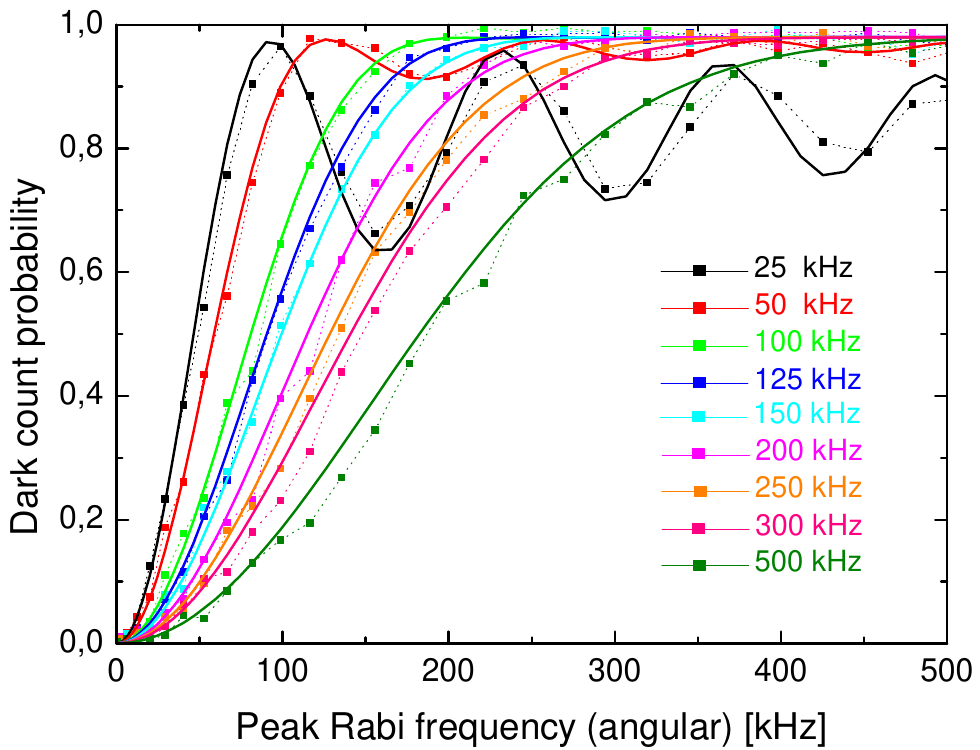}
\caption[RAP efficiency versus peak Rabi frequency for different chirp ranges]{(color online) RAP efficiency versus peak Rabi frequency for different chirp ranges: The plot shows resulting transfer efficiencies while the peak amplitude is scanned. The solid lines are obtained from a numerical solution of the time-dependent Sch\"odinger equation for the qubit, including thermal effects by averaging over a distribution Rabi frequencies (see text). For small chirp ranges, one observes a Rabi oscillation-like behavior, while adiabaticity is attained for chirp ranges of 100~kHz or larger. For even larger chirp ranges, the increase in robustness is bought at the expense of a higher power requirement. Note that no free parameters were used for the simulation, all parameters were inferred from the pulse width scan measurement of Fig. \ref{fig:quadcohdynamics} and the power gauge measurement.}
\label{fig:rapdifferentchirprange}
\end{center}\end{figure}
For the detailed evaluation of the RAP pulses, we first gauged the bare Rabi frequency $\Omega_0$ versus the precisely controllable r.f. voltage which is supplied to the AOM. The maximum excitation time is measured by recording the first oscillation period of a signal such as the one shown in Fig. \ref{fig:quadcohdynamics} for a set of different r.f. amplitudes. $\Omega_0$ is then determined by extracting the effective Rabi frequency $\bar{\Omega}$, as explained above, and making use of Eq. \ref{eq:effectivepitime}. The resulting Rabi frequencies are used for a fit to a third order polynomial function, such that the Rabi frequency for any r.f. amplitude can be obtained. In the experiment, we use RAP pulses with temporally varying Rabi frequency $\Omega(t)$ and detuning $\delta(t)$ according to
\begin{eqnarray}
\Omega(t)&=&\Omega_0^{\textrm{(cal)}}e^{-\frac{t^2}{2\tau_{\sigma}^2}} \nonumber \\ 
\delta(t)&=&\pi r_c t/\tau_{\sigma},
\label{eq:chirppulse}
\end{eqnarray}
where $\Omega_0^{\textrm{(cal)}}$ the peak amplitude known from the power calibration, $\tau_{\sigma}$ determines the duration of the pulse, which has a Gaussian envelope, and the linear frequency chirp across the resonance is determined by the chirp range $r_c$. The pulses are truncated in time at $t\pm 2\tau_{\sigma}$, and both frequency and amplitude are changed in 50 discrete time steps. Fig. \ref{fig:rapdifferentchirprange} shows the excitation probability, i.e. the probability to find the ion in the metastable state, versus peak Rabi frequency for different chirp ranges and fixed $\tau_{\sigma}=$50~$\mu$s, resulting in a 4~$\mu$s sample duration. It is important to note that a phase-continuous frequency switching mode of the r.f. synthesizer is used, where the phase of each r.f. sample is chosen such that the output signal is continuous. One clearly observes the transition from a Rabi-oscillation behavior to a monotonic increase in excitation with the amplitude as the chirp range becomes larger and adiabatic following conditions are attained at $r_c=$100~kHz. For larger chirp ranges, higher amplitudes are needed to attain maximum excitation, as the effective time on resonance is decreased. Fig. \ref{fig:rapdifferentchirprange} also displays the results of simulations for each chirp range, where the time-dependent Schr{\"o}dinger equation is solved numerically for the excitation pulse Eq. \ref{eq:chirppulse}, while the averaging over the thermal Rabi frequency distribution Eq. \ref{eq:normalizedmodeledeffectiverabifreqdist} is done. The parameter $b=$7.1$\cdot$10$^{-4}$ is chosen according to measurement results from Fig. \ref{fig:quadcohdynamics}, such that no free fit parameter is occurring.

\begin{figure}[t!]\begin{center}
\includegraphics[width=0.45\textwidth]{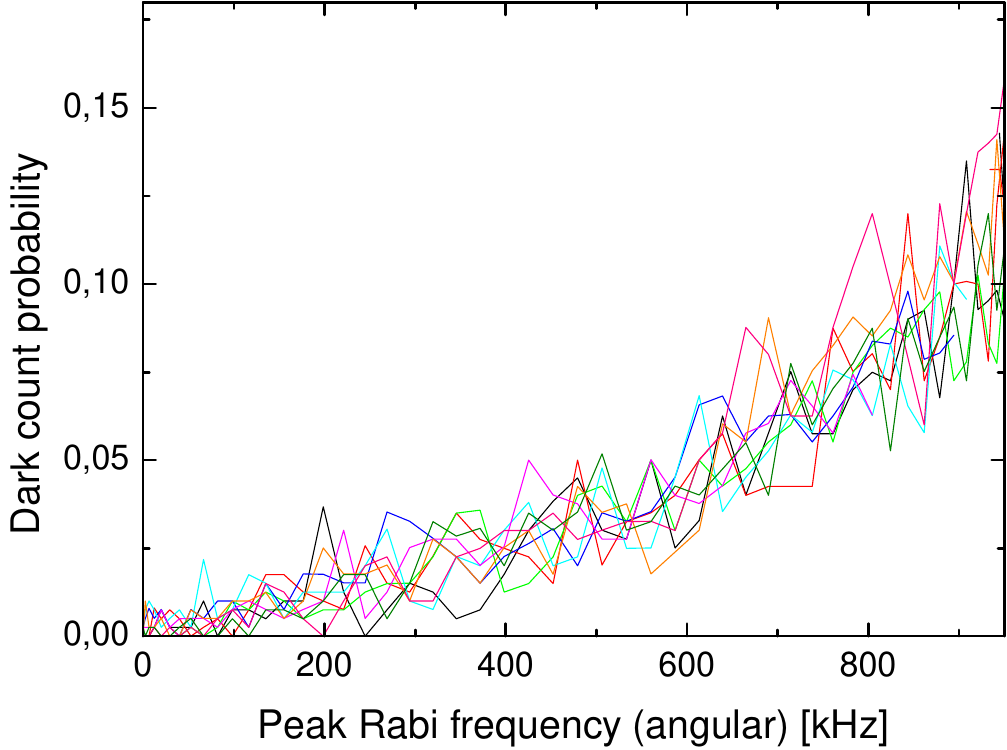}
\caption[Parasitic shelving]{(color online) Parasitic shelving: The figure shows the shelved population versus peak Rabi frequency for the same parameters as in Fig. \ref{fig:rapdifferentchirprange}. The central frequency is resonant with the $\ket{\uparrow}\rightarrow \ket{D_{5/2},m_j=+5/2}$ transition, but the ion was initialized in $\ket{\downarrow}$ such that no population should be transferred under ideal conditions. Data for different chirp ranges is shown. It can be seen that population transfer is insensitive to the chirp range, which suggests a completely incoherent transfer mechanism.}
\label{fig:shelvingfromdw}
\end{center}\end{figure}

The fidelity of readout process can also be deteriorated if population is excited from the \textit{wrong} qubit state, i.e. $\ket{\downarrow}$ in our case. Conceivable mechanisms for this transfer are off-resonant excitation on parasitic transitions, resonant excitation of motional sidebands of these transitions, or resonant excitations by frequency components of the excitations pulse which arise from the discrete sampling. Fig. \ref{fig:shelvingfromdw} shows the population transferred to the metastable D$_{5/2}$ state by similar pulses as in Fig. \ref{fig:rapdifferentchirprange}, but for initialization in the $\ket{\downarrow}$ level. One can see that an error of about 1\% is present at a peak Rabi frequency of 200~kHz, and the excitation is independent of the chirp range. The transition connecting to $\ket{\downarrow}$ with the lowest frequency offset is the $\ket{\downarrow}\rightarrow\ket{D_{5/2},m_J+3/2}$ transition, which is separated from the driven $\ket{\uparrow}\rightarrow\ket{D_{5/2},m_J=+5/2}$ by about 8~MHz. This  is far beyond the used chirp ranges, i.e. out of the bandwidth of the excitation pulses. The simulation yields only population transfer values up to 10$^{-5}$ for the maximum amplitude, where the values are strongly increasing with the chirp range. Thus, the excitation mechanisms via resonant frequency components and resonant excitation of motional sidebands can therefore be excluded. It can therefore be concluded that the parasitic transition is excited by incoherent amplified spontaneous emission (ASE) background from the amplified laser. Thus, this is a purely technical artifact which can be overcome by employing a filter cavity.

\section{Numerical study of fidelity and robustness}
\label{sec:numerical}
\begin{figure*}[th!]\begin{center}
\includegraphics[width=0.9\textwidth]{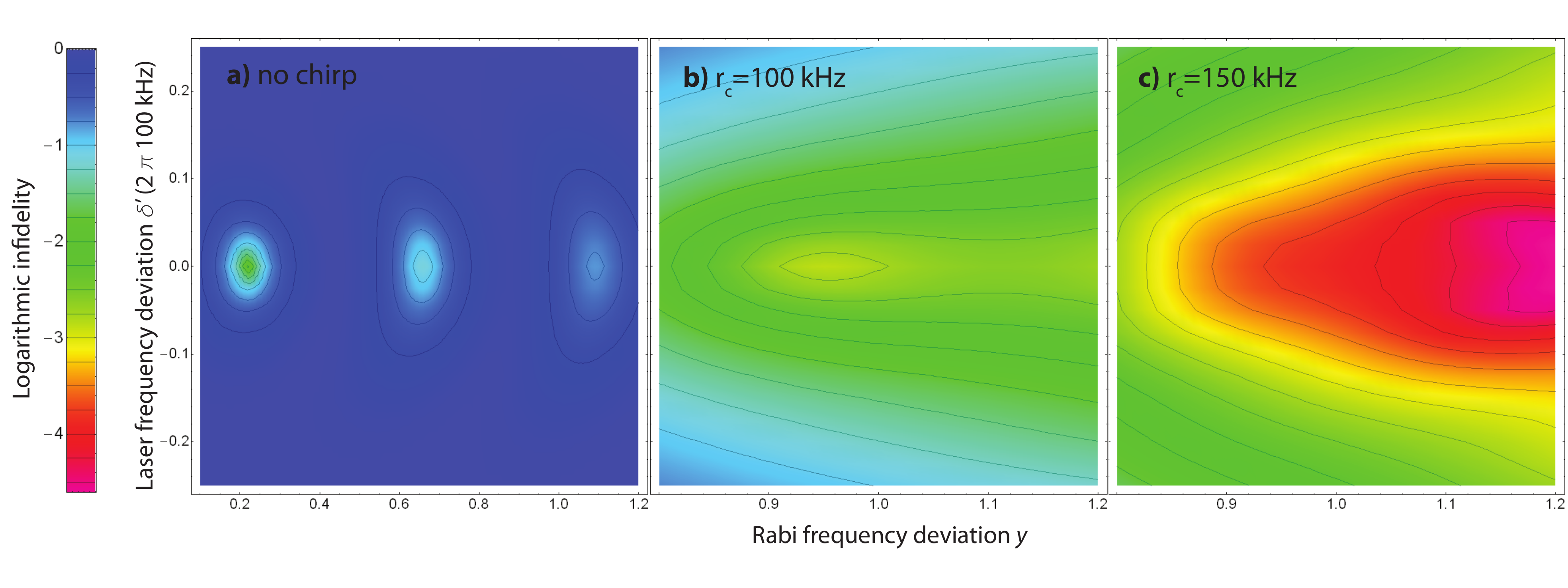}
\caption[RAP efficiency versus peak Rabi frequency for different chirp ranges]{(color online) Simulation of the RAP pulse effect for systematic deviation of experimental parameters. The plots show simulation data of the logarithmic infidelity $\log_{10}\left(1-p_D(t_e)\right)$, i.e. a measure for the population not transferred to the metastable state. The infidelities are plotted versus a factor describing static errors of the Rabi frequency, $y=\Omega_0/\Omega_0^{\textrm{(cal)}}$ with respect to the one obtained from the calibration $\Omega_0^{\textrm{(cal)}}$, and versus an extra detuning $\delta'$ describing an error of the laser frequency with respect to the atomic resonance, which is given in units of the chirp range of $2\pi\cdot$100~kHz from plot b). Plot \textbf{a)} shows the infidelity data for zero chirp range, i.e. in the Rabi oscillation regime. The peak Rabi frequency is chosen to be $\Omega_0^{\textrm{(cal)}}\approx 2\pi\cdot$332~kHz. One can clearly see that the parameter regions for which a considerably high population transfer is accomplished are rather low, and a minimum infidelity of about -2 is attained. Plot \textbf{b)} shows the infidelity for a chirp range of $2\pi\cdot$100~kHz and a peak Rabi frequency of $\Omega_0^{\textrm{(cal)}}\approx 2\pi\cdot$221~kHz, corresponding to the situation where adiabatic following conditions are fulfilled and the transfer efficiency saturates with respect to laser power as inferred from the experimental data Fig. \ref{fig:rapdifferentchirprange}. Plot \textbf{c)} shows the same simulation for a chirp range of $2\pi\cdot$150~kHz and a correspondingly increased Rabi frequency of $\Omega_0^{\textrm{(cal)}}\approx 2\pi\cdot$332~kHz.
}
\label{fig:robustness}
\end{center}\end{figure*}
The RAP pulses are characterized by the parameters duration, amplitude and chirp range, and a quantitative understanding of the role of these parameters is required to achieve predetermined fidelity and robustness values. As the situation is complicated due to the presence of thermal fluctuations, the Maxwell-Boltzmann model proves to be a useful tool for a precise numerical characterization of the dynamics. For a given thermal Rabi frequency reduction factor $x$, we numerically solve the Schr\"odinger equation for the qubit in ODE form:
\begin{eqnarray}
\dot{c}_g^{(x)}&=&\tfrac{i}{2}\left((\delta(t)+\delta')c_g^{(x)}+xy\Omega(t)c_e^{(x)}\right) \nonumber \\
\dot{c}_e^{(x)}&=&\tfrac{i}{2}\left(-(\delta(t)+\delta')c_e^{(x)}+xy\Omega(t)c_g^{(x)}\right).
\label{eq:tlsode}
\end{eqnarray}
Here, $c_g(t)$ and $c_e(t)$ are the amplitudes of the ground and excited state, respectively, and the system is assumed to be perfectly prepared in the electronic ground state prior to the excitation pulse. $\delta'$ represents an additional offset detuning arising from possible drifts of the laser frequency, and $y$ is an additional constant scaling factor for the Rabi frequency from laser intensity drifts. For a given parameter set $\delta'$,$y$, Eqs. \ref{eq:tlsode} are numerically solved for $0\leq x\leq 1$ in steps of $\Delta x=0.01$ and the resulting transfer efficiency $p_D^{(x)}(t_e)=|c_e^{(x)}(t_e)|^2$ after the pulse is averaged according to
\begin{equation} 
p_D(t_e)=\sum_x \Delta x\; w_b(x\Omega_0) p_D^{(x)}(t_e)
\end{equation}
with $w_b(\Omega)$ from Eq. \ref{eq:normalizedmodeledeffectiverabifreqdist}. \\
Fig. \ref{fig:robustness} shows the resulting logarithmic infidelity, i.e. the population residing in the initial state after the transfer pulse, versus the frequency and amplitude error parameters for three different chirp ranges. The duration of $\tau_{\sigma}=$50~$\mu$s, the $b$ parameter and the sample number are chosen to be constant and  identical to the experimental parameters from Fig. \ref{fig:rapdifferentchirprange}. Dephasing Rabi oscillation behavior is seen for zero chirp range, and it becomes clear that the frequency and amplitude have to be rather precisely calibrated and stabilized in order to keep the transfer efficiency sufficiently high. By contrast, chirp ranges of 100~kHz and more lead to a very robust behavior. For an increased chirp range of 150~kHz, extra laser power is needed to keep adiabatic following conditions satisfied, but the maximum transfer efficiency is increased by more than one order of magnitude, and the robustness window is greatly enhanced.  This illustrates how the RAP process allows to make use of extra laser power to obtain an increased robustness and a higher maximum efficiency. Note that transfer infidelities below 0.1\% can hardly be actually observed in the experiment as other limitations from the state preparation and fluorescence readout steps obscure the result.

\section{Conclusion and Outlook}
We have developed a Maxwell-Boltzmann theory for thermally induced Rabi frequency fluctuations in the coherent excitation of a single trapped ion with a laser. The model is successfully applied to reproduce the dephasing behavior of Rabi oscillations and to RAP excitation pulses with variable parameters. The latter opens the possibility to make the right choice of pulse parameters to achieve well defined behavior in terms of robustness and efficiency if the ion temperature and the relation between Rabi frequency and laser power is known. Based on the model, we have carried out numerical simulations which allow for the determination of the robustness against experimental parameter fluctuations. This makes the RAP pulse a very reliable tool for population transfer in quantum experiments with trapped ions. Furthermore, the model offers a new thermometry scheme which operates from closely above the Lamb-Dicke regime to temperatures several times larger than the Doppler cooling limit temperature. It closes a gap between two established methods: The fully coherent measurement method which compares the Rabi frequencies of the red and blue motional sidebands on a given motional mode, which works only in the Lamb-Dicke regime and requires that either the interaction with spectator modes is suppressed or that these are also cooled deeply into the Lamb-Dicke regime. A simpler scheme based on time-resolved fluorescence readout \cite{EPSTEIN2007,WESENBER2007} does not require the ability of coherently driving long-lived transitions, but ceases to yield precise answers at temperatures close to and below the Doppler limit.\\
Our technique also opens the possibility for the application of quantum control strategies such as optimal control theory for thermal ensembles, e.g. for improving the robustness of quantum gates. Entangling gates are usually the operations in quantum experiments with trapped ions which are the most difficult to realize with considerable fidelity, as the interplay between internal and external degrees of freedom requires the precise knowledge and control of many parameters. Residual thermal excitation of the motional modes leads to a loss of controllability  of the system \cite{POSCHINGER2010}. A recent example of how coherent control strategies can cure the effect of control parameter fluctuations for an entangling quantum gate is given by \cite{HAYES2011}. The averaging of the coherent dynamics as in Eq. \ref{eq:effectivemoderabiosc} for general Hamiltonians in conjunction with optimal control theory \cite{NEBENDAHL2009} could lead to the development of gate schemes which are more robust in the thermal regime than current ones \cite{KIRCHMAIR2009}.

\vspace{1cm}

We acknowledge financial support by the European commission within the IP AQUTE and by IARPA with the SQIP program.

% Bibliography
\bibliographystyle{unsrt}

%\bibliography{./Ref_lib_ions}

\begin{thebibliography}{10}

\bibitem{TIMONEY2008}
N.~Timoney, V.~Elman, S.~Glaser, C.~Wei{\ss}, M.~Johanning, W.~Neuhauser, and
  Chr. Wunderlich.
\newblock Error-resistant single qubit gates with trapped ions.
\newblock {\em Phys. Rev. A}, 77:052334, 2008.

\bibitem{KIRCHMAIR2009}
G.~Kirchmair, J.~Benhelm, F.~Z{\"a}hringer, R.~Gerritsma, C.~F. Roos, and
  R.~Blatt.
\newblock Deterministic entanglement of ions in thermal states of motion.
\newblock {\em New. J. Phys.}, 11:023002, 2009.

\bibitem{POSCHINGER2009}
U.~G. Poschinger, G.~Huber, F.~Ziesel, M.~Deiss, M.~Hettrich, S.~A. Schulz,
  G.~Poulsen, M.~Drewsen, R.~J. Hendricks, K.~Singer, and F.~Schmidt-Kaler.
\newblock Coherent manipulation of a $^{40}$ca$^+$ spin qubit in a micro ion
  trap.
\newblock {\em J. Phys. B: At. Mol. Opt. Phys.}, 42:154013, 2009.

\bibitem{KREUTER2005}
A.~Kreuter, C.~Becher, G.P.T. Lancaster, A.B. Mundt, C.~Russo, H.~H{\"a}ffner,
  C.~Roos, W.~H{\"a}nsel, F.~Schmidt-Kaler, and R.~Blatt.
\newblock Experimental and theoretical study of the 3$d^{2}d$-level lifetimes
  of $^{40}$ca$^{+}$.
\newblock {\em Phys. Rev. A}, 71:032504, 2005.

\bibitem{WUNDERLICH2007}
Chr. Wunderlich, Th. Hannemann, T.~Koerber, H.~Haeffner, Ch. Roos, W.~Haensel,
  R.~Blatt, and F.~Schmidt-Kaler.
\newblock Robust state preparation of a single trapped ion by adiabatic
  passage.
\newblock {\em Journal of Modern Optics}, 54:1541, 2007.

\bibitem{LEIBFRIED2003}
D.~Leibfried, R.~Blatt, C.~Monroe, and D.J. Wineland.
\newblock Quantum dynamics of single trapped ions.
\newblock {\em Rev. Mod. Phys.}, 75:281, 2003.

\bibitem{ROOS2000}
C.~Roos.
\newblock {\em Controlling the quantum state of trapped ions}.
\newblock PhD thesis, Leopold-Franzens-Universit{\"a}t Innsbruck, p. 117, 2000.

\bibitem{SCHULZ2006}
S.~Schulz, U.~Poschinger, K.~Singer, and F.~Schmidt-Kaler.
\newblock Optimization of segmented linear paul traps and transport of stored
  particles.
\newblock {\em Fortschr. Phys.}, 54:648, 2006.

\bibitem{SCHULZ2008}
S.~Schulz, U.~Poschinger, F.~Ziesel, and F.~Schmidt-Kaler.
\newblock Sideband cooling and coherent dynamics in a microchip multi-segmented
  ion trap.
\newblock {\em New J. Phys.}, 10:045007, 2008.

\bibitem{EPSTEIN2007}
R.~J. Epstein, S.~Seidelin, D.~Leibfried, J.~H. Wesenberg, J.~J. Bollinger,
  J.~M. Amini, R.~B. Blakestad, J.~Britton, J.~P. Home, W.~M. Itano, J.~D.
  Jost, E.~Knill, C.~Langer, R.~Ozeri, N.~Shiga, and D.~J. Wineland.
\newblock Simplified motional heating rate measurements of trapped ions.
\newblock {\em Phys. Rev. A}, 76:033411, 2007.

\bibitem{WESENBER2007}
J.~H. Wesenberg, R.~J. Epstein, D.~Leibfried, R.~B. Blakestad, J.~Britton,
  J.~P. Home, W.~M. Itano, J.~D. Jost, E.~Knill, C.~Langer, R.~Ozeri,
  S.~Seidelin, and D.~J. Wineland.
\newblock Fluorescence during doppler cooling of a single trapped atom.
\newblock {\em Phys. Rev. A}, 76:053416, 2007.

\bibitem{POSCHINGER2010}
U.~G. Poschinger, A.~Walther, K.~Singer, and F.~Schmidt-Kaler.
\newblock Observing the phase space trajectory of an entangled matter wave
  packet.
\newblock {\em Phys. Rev. Lett.}, 105:263602, 2011.

\bibitem{HAYES2011}
D.~Hayes, S.~M. Clark, S.~Debnath, D.~Hucul, Q.~Quraishi, and C.~Monroe.
\newblock Coherent error suppression in spin-dependent force quantum gates.
\newblock {\em arXiv:quant-ph/1104.1347}, 2011.

\bibitem{NEBENDAHL2009}
V.~Nebendahl, H.~H{\"a}ffner, and C.~F. Roos.
\newblock Optimal control of entangling operations for trapped-ion quantum
  computing.
\newblock {\em Phys. Rev. A}, 79:012312, 2009.

\end{thebibliography}

\end{document}